\documentclass[%
reprint,
amsmath,amssymb,  aps, prc, nofootinbib
]{revtex4-1}

\usepackage{graphicx}% Include figure files
\usepackage{dcolumn}% Align table columns on decimal point
\usepackage{bm}% bold math
\begin{document}

\title{Jet quenching in the hadron gas: an exploratory study}

\author{P.~Dorau$^{1}$, J.-B.~Rose$^{1,2}$, D.~Pablos$^{3}$, and H.~Elfner$^{4,1,2}$}

\affiliation{$^1$Institute for Theoretical Physics, Goethe University,
Max-von-Laue-Strasse 1, 60438 Frankfurt am Main, Germany}
\affiliation{$^2$Frankfurt Institute for Advanced Studies, Ruth-Moufang-Strasse 1, 60438
Frankfurt am Main, Germany}
\affiliation{$^3$Institutt for fysikk og teknologi, University of Bergen, Postboks 7803, 5020 Bergen, Norway}
\affiliation{$^4$GSI Helmholtzzentrum f\"ur Schwerionenforschung, Planckstr. 1, 64291
Darmstadt, Germany}

\date{\today}

\begin{abstract}
The suppression of high momentum particles in heavy-ion collisions in comparison to elementary reactions is one of the main indications for the formation of a quark-gluon plasma. In recent studies, full jets are being reconstructed and substructure observables are gaining importance in assessing the medium modifications of hard probes. In this work, the effect of the late stage hadronic interactions are explored within the hadronic transport approach SMASH (Simulating Many Accelerated Strongly-interacting Hadrons). High momentum particles are incorporated in a radially expanding hadron gas to analyse the corresponding angular distributions, also refered to as `jet shape' observables. We find that the full hadron gas can be approximated with a pion gas with constant elastic cross-sections of 100 mb. In addition, the temperature and probe energy dependence of diffusion coefficients $\tilde{q}$ and $\tilde{e}$ quantifying the transverse and parallel momentum transfers are extracted. The species dependence and the importance of different interaction types are investigated. Parametrizations are presented that can be employed in future jet quenching calculations to include the effect of the hadronic phase.  
\end{abstract}

\maketitle

\section{\label{intro}Introduction}

The study of the strong suppression and substructure modification of high energy jets created in heavy-ion collisions offers the opportunity to characterise the properties of the medium they traversed. These phenomena, typically referred to as jet quenching \cite{Mehtar-Tani:2013pia,Qin:2015srf}, have so far been analysed mainly by modelling the interaction of colored partons with the deconfined, high temperature phase of the medium, the strongly coupled liquid known as the quark-gluon plasma (QGP) (for a recent review on the main open questions about our current understanding of QGP see \cite{Busza:2018rrf}). It is well known that at lower temperatures the fluid experiences a process of `particlization' and the system is well described by the dynamics of a hadron resonance gas \cite{Petersen:2014yqa}. Nevertheless, the vast majority of studies that address jet quenching phenomenology have so far omitted, without any compelling quantitative justification, the effects of the interaction of the hadronic components of the jet with this lower temperature phase of the medium. 
%With the present exploratory study we provide benchmarks of the magnitude and features of such interactions in order to guide and motivate their inclusion in jet quenching models in the immediate future.

Jet quenching physics is a very active field within that of heavy-ion collisions, partly due to the wealth of experimental jet data that has been produced in recent times at RHIC and LHC. Jets are sprays of hadrons which are clustered together according to a specific reconstruction algorithm with a given jet radius parameter $R$. They are the result of the fragmentation of parton showers that are developed through the relaxation of the virtuality scale $Q^2 \sim \mathcal{O}(p_{T, \rm{jet}}^2)$ via successive splittings down to the hadronization scale. In heavy-ion collisions, the dynamically evolving parton shower interacts with the strongly coupled QGP, which has a temperature $T$ of the order of the non-perturbative scale $\Lambda_{\rm QCD}$. The complicated nature of this multi-scale, multi-partonic problem makes it difficult to achieve a unified, self-consistent picture of jet quenching physics, and for this reason several theoretical descriptions of the jet/medium interplay have been pursued, both from a perturbative (see \cite{Mehtar-Tani:2013pia,Qin:2015srf} and references therein) and non-perturbative (see \cite{Chesler:2015lsa} and references therein) perspective. Each of these descriptions inherently makes assumptions on the nature of the relevant degrees of freedom of the QGP to which an energetic jet is sensitive to, and thus allow for the falsification of different pictures of the inner workings of the plasma through the comparison between models and experimental data. However, without the inclusion of the potentially important observable effects due to the hadron gas phase, any conclusion drawn from such phenomenological studies cannot be definitive.

In \cite{Cassing:2003sb} the nuclear modification factor at RHIC has been studied within a purely hadronic approach achieving a qualitative description of the experimental data including a significant suppression. More recently similar studies are carried out within Angantyr+UrQMD at LHC energies and confirm significant suppression effects within the hadronic stage \cite{Bierlich:2018xfw}. More complete dynamical approaches including soft and hard contributions in the same approach like \cite{Werner:2012xh,Werner:2012sv} show that qualitative differences in the nuclear modification factor and elliptic flow are expected in the intermediate momentum region from $p_{\perp} = 2-6$ GeV. 

The precise way in which the products of the fragmentation of a parton shower start being sensitive to the presence of the hadronic medium is, in fact, not so well understood. This question involves both the space-time picture associated to the formation of the parton shower as well as the time associated to the process of hadronization itself, which could in turn result from the recombination of partons from the jet with those from the thermal medium \cite{Fries:2003vb,Han:2016uhh}. Studying the implications of this modelling is beyond the scope of this paper and will be left for future work. The goal of the current work consists instead in quantifying the effects of the hadronic medium on energetic hadrons in a controlled scenario. In this way, we are able to provide the first solid indications of the importance of quenching in the hadronic gas phase through the analysis of angular energy distributions, which is an observable analogous to those usually used in jet quenching phenomenology to characterise how energy is spread away from the jet axis due to medium interactions \cite{Chatrchyan:2013kwa,Aad:2019igg,Acharya:2019ssy}. We further parametrise the magnitude of the effect in terms of the newly defined $\tilde q$ and $\tilde e$, the average transverse momentum and longitudinal momentum transferred per unit of mean free path, respectively. Even though their size will be shown to be around 3 to 4 times smaller than their QGP counterparts, namely $\hat{q}$ and $\hat{e}$, there is no reason why they should be neglected, and therefore their impact on jet observables, especially on jet substructure observables, is predicted to be sizeable.

The rest of the paper is organised as follows: in Section~\ref{smash} the hadronic transport model SMASH used in this work is presented. Then, in Section~\ref{shapes} we show results on the angular energy distributions for incoming hadrons with different energies and species traversing an expanding hadronic medium of different lengths. In Section~\ref{qhat} we parametrise the relevant quantities $\tilde q$ and $\tilde e$ in terms of momentum and temperature by carrying out a detailed microscopic analysis for individual collisions. Finally, in Section~\ref{conclusion} we summarise our findings and discuss their potential relevance in the understanding of jet quenching observables.

\section{\label{smash}Hadronic transport: SMASH}

In this work we use SMASH v1.6 \cite{Weil:2016zrk,SMASHwebsite,SMASH1.6DOI}, a newly developed transport approach, to describe the hadronic medium. This model has been shown \cite{Tindall:2016try} to effectively solve the Boltzmann equation
\begin{equation} \label{eq:boltzmann}
 p^\mu \frac{ \partial f_i (t,{\bf x},{\bf p})}{\partial x^\mu}  = C_{coll} [f_i,f_j] \ ,
\end{equation}
where $f_i$ is a one-particle distribution function for species $i$ and $C_{coll}$ is the collision integral describing the interaction of the various species.

\begin{figure}
 \includegraphics[width=85mm]{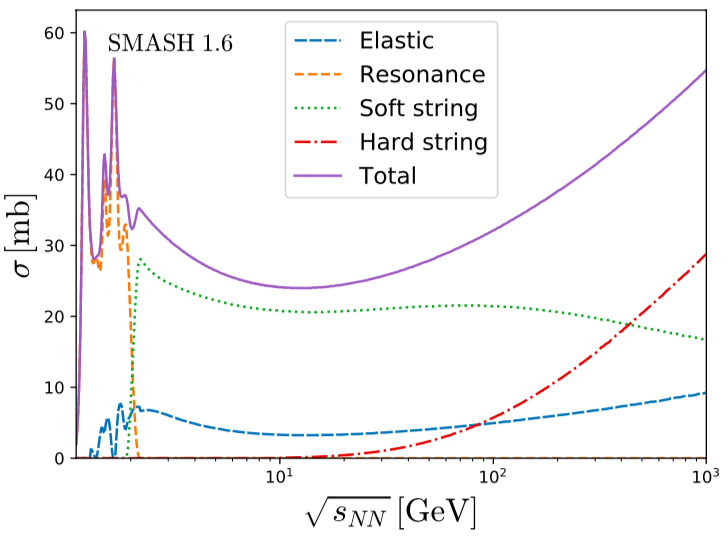}
\caption{Total $\pi^- p$ cross-section in SMASH, decomposed according to the various partial contributions. Taken from \cite{Mohs:2019iee}}
\label{pip_xsec}
\end{figure}

Particles in SMASH propagate according to their equations of motion (no potentials are used in this work, so this corresponds to straight lines), and collide according to a geometric criterion
\begin{equation}
d_{trans} \le \sqrt{\frac{\sigma_{tot}}{\pi}},
\label{eq:coll_crit}
\end{equation}
with $d_{trans}$ the distance between two particles and $\sigma_{tot}$ their total cross-section, which depends on the incoming particle species and momenta. Degrees of freedom include all confirmed particles from the PDG 2018 list \cite{Tanabashi:2018oca} up to masses of $\sim$2.3 GeV. As can be seen for example in the $\pi^-$-proton cross-section (Fig.~\ref{pip_xsec}), at low energies inelastic interactions happen through resonance formation, and after a transition region, the high energy regime uses a string excitation and fragmentation model. Soft strings are excited directly within SMASH according to single-diffractive direct excitation, double-diffractive gluon exchange or non-diffractive quark exchange, and subsequently fragmented using PYTHIA 8.235 \cite{Andersson:1983ia, Sjostrand:2014zea}; hard strings are both excited and fragmented directly within PYTHIA (see \cite{Mohs:2019iee} for the full information on the treatment of strings in SMASH).

Note that although some cross-sections such as the $\pi^- p$ shown in Fig.~\ref{pip_xsec} are quite well known experimentally, SMASH features a very high number of particle pairs for which experimental cross-section data is scarce or altogether inexistent. Thus we use the Additive Quark Model (AQM) \cite{Goulianos:1982vk} as a prescription for the prediction of unknown cross-sections. In the AQM, the high-energy cross-section of species $i$ and $j$ can be scaled from a known elastic or total cross-section, for example such that
\begin{equation}
\sigma_{ij} = \frac{\sigma_{ij}^{AQM}}{\sigma_{\pi p}^{AQM}} \sigma_{\pi p},
\end{equation}
where
\begin{equation}
\sigma^{AQM} = 40 \cdot \left( \frac{2}{3} \right)^{n_M} \cdot (1-0.4x^s_1) \cdot (1-0.4x^s_2).
\label{sigma_aqm}
\end{equation}
In this last formula, $n_M$ is the number of incoming mesons in the reaction, and $x^s$ is the fraction of strange quarks in the considered hadron. In SMASH, baryon-baryon interactions follow the nucleon-nucleon cross-section, while baryon-meson and meson-meson interactions follow the pion-nucleon cross-section.

In the following sections we will consider two initialization schemes for the study of high-$p_T$ particles within SMASH. The first one, used in Section \ref{shapes}, is a uniformly dense sphere of radius $r$ initialized with thermal multiplicities and momenta. This sphere expands over time and eventually freezes out as the density decreases, reaching a state of outwards free-streaming particles at large times. The second initialization, used in Section \ref{qhat} corresponds to an infinite medium in the form of a box with periodic boundary conditions, and is uniformly filled in the same way as the sphere. This system conserves density and thermodynamic quantities such as temperature over the volume for longer times, provided detailed balance is enforced\footnote{This is not strictly the case if strings are enabled (which is necessary for the current study to be sensible) leading to a gradual decrease in the temperature. Since we are however only interested in what happens at relatively early times (typically less than 10-20 fm), we will simply neglect this effect (over such times, the temperature typically will not decrease by more than 3\%).}.

\section{\label{shapes}Jet Shapes}

\begin{figure*}
\centering
 \includegraphics[width=85mm]{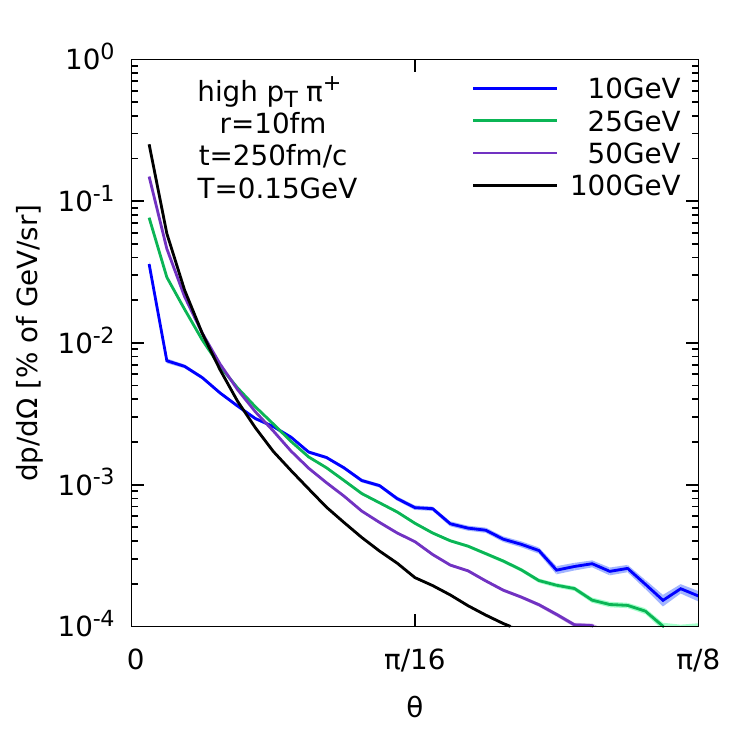}
 \includegraphics[width=85mm]{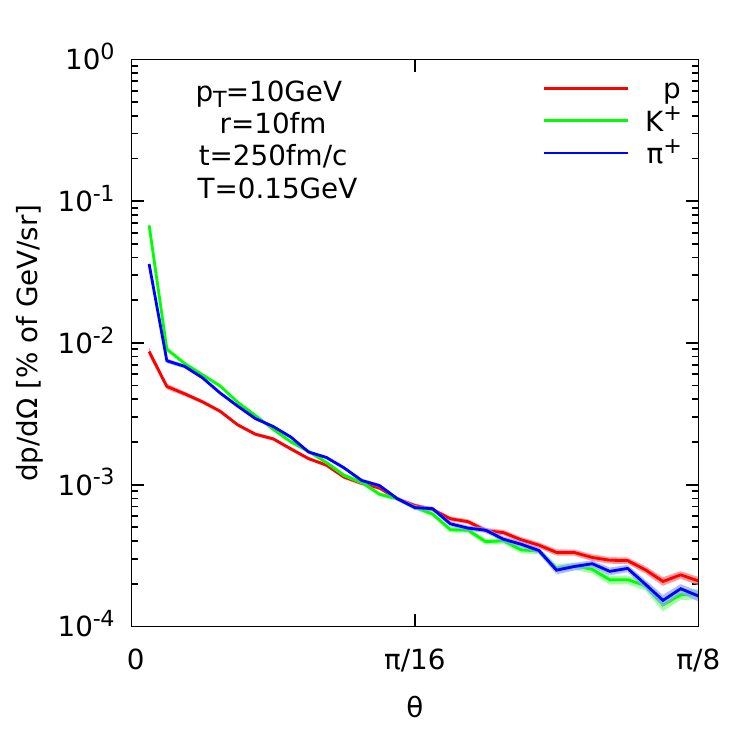}
 \includegraphics[width=85mm]{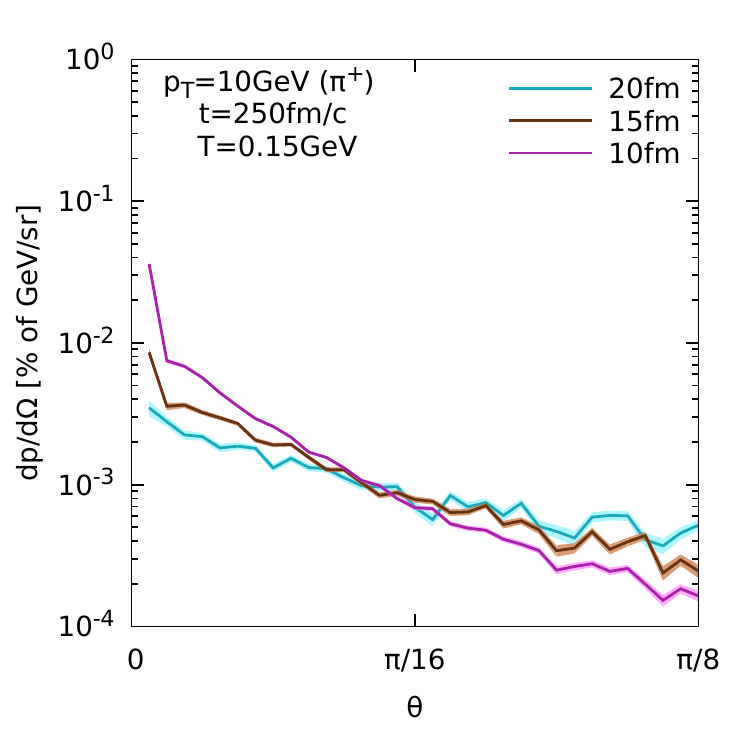}
 \includegraphics[width=85mm]{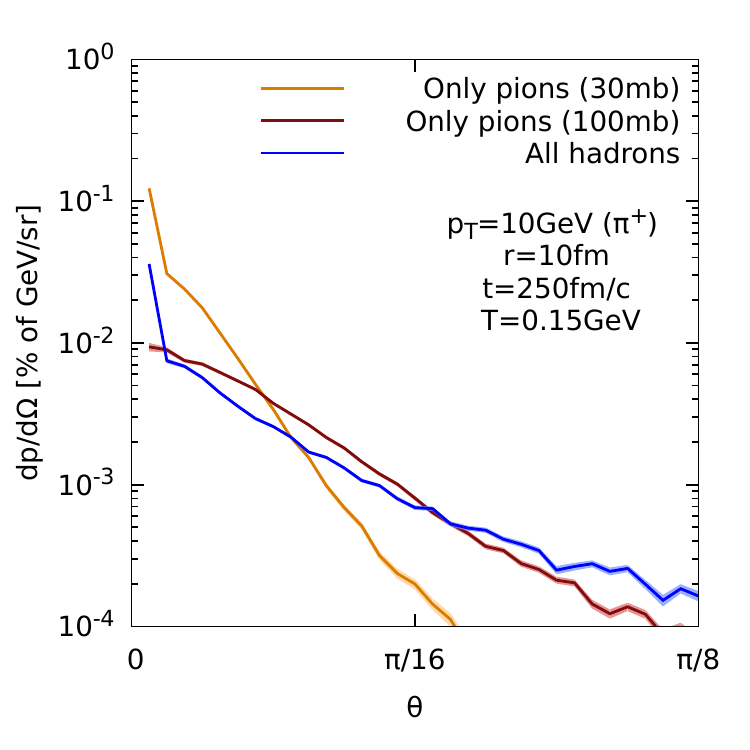}
\caption{Jet shapes as a function of leading particle energy (top left), leading particle species (top right), medium size (bottom left). The bottom right plot compares the full hadron gas jet shape to a pion gas with constant cross-section jet shape.}
\label{jetshapes}
\end{figure*}

In this section we add a high-$p_T$ particle in the middle of the previously described thermally initialized sphere at a temperature of $T=150$ MeV, and measure the angular distribution at large times, after freeze-out. To some extent, one can consider this scenario as similar to what would happen to a high-$p_T$ particle in the late stages of a heavy ion collision, after the hadronization has taken place: at that point, it crosses a rapidly cooling and decreasingly dense hadronic medium.

We determine the so-called "jet shapes" shown in Fig.~\ref{jetshapes} by measuring the amount of energy such a high-$p_T$ particle adds on average at an angle $\theta$ of its original propagation direction. Specifically, this is done by simulating both a set of spheres in which this high-$p_T$ particle is present and and one in which it is absent. At a given angle $\theta$ corresponding to a solid angle $\Omega$, the momentum $dp/d\Omega$ of the latter is subtracted from the momentum of the first, in what can be thought of as a background subtraction. We then normalize this by dividing it by the maximum value it could take, i.e. if all the momentum was still at $\theta=0$ (which corresponds to the case where the particle flies out of the medium without interacting). This gives us a quantity which can be compared for a wide variety of scenarios.

The top left panel of Fig.~\ref{jetshapes} explores the energy dependence of the high-$p_T$ particle. As one can readily see, less energetic particles tend to affect the momentum distribution at wider angles than their more energetic counterparts. This is expected, as we would indeed expect that a 100 GeV particle, even if it does interact with the medium, should retain or transfer most of its momentum in the initial direction of propagation when colliding elastically or inelastically with a medium component with energy of the order of 1-3 GeV. Conversely, using the same considerations, a much larger part of the momentum goes to wider angles in the case of a 10 GeV particle. The peak around zero angle, e.g. for the 10 GeV probe, reflects the particles that escape from the medium without any disturbance. 

We see in the top right panel of Fig.~\ref{jetshapes} that different species of particles are differently affected by the hadronic expansion. This is due to different hadrons having on average larger or smaller cross-sections with the particles of the medium. Remembering Eq.~\eqref{sigma_aqm}, we then see that the proton, as a baryon, typically has larger cross-section with the medium, and its jet shape is more skewed towards larger angles; the pion shape is less affected, but still slightly more so than the strange kaon one, with the smallest average cross-section.

The medium size dependence (here probed through varying the radius of the initial expanding sphere) is inspected in the bottom left panel of Fig.~\ref{jetshapes}. As one would expect, increasing the size of the medium (and thus the number of possible collisions between the high-$p_t$ and medium particles) generally broadens the angular distribution. Note that while the size of the hadronic part of the medium in heavy ion collisions is not precisely known, estimations usually place it between 10 and 15 fm \cite{Petersen:2008gy}.

Finally, the bottom right panel compares shooting a high-$p_T$ pion through a full hadron gas as described in the previous section, and through a much simpler pion gas interacting only through constant cross-sections (this corresponds in essence to the hard spheres scenario). Although the 30mb case is much closer to the actual average cross-section the pion would encounter in a full hadron gas, we see that the angular distribution of the 100 mb case is in fact much closer to that of the full hadron gas due to the much larger density than in the pure pion gas. In order to get an intuition of the degree at which the 30 mb and 100 mb cases differ, we calculate the proportion of volume at thermal densities which is occupied by particles in this hard sphere scenario. For an initial temperature of 150 MeV, we thus see that $\sim 5\%$ of the volume is occupied in the case of the 30 mb cross-sections, whereas $\sim 33 \%$ of space is filled in the case of the 100 mb cross-sections. Although this is a simplified model, it should provide the reader with some sense of how dense such a hadron gas really is at the time of hadronization.

This explorative investigation shows that the angular distributions can be affected by the hadronic phase rescatterings in a significant manner. In particular, let us note that fully reconstructed jets rely on information of particles at much lower transverse momenta as well. Since the probe hadron would be of similar energy as the medium particles, making it hard to distinguish it clearly, in our radially expanding sphere it does not make sense to lower the momenta beyond the displayed 10 GeV. However hadrons around 2 GeV of energy would certainly be re-shuffled and found at different angles than without hadronic rescattering, sizeably affecting the tail of the distributions of observables such as the `jet shapes' computed in \cite{Chien:2015hda,Casalderrey-Solana:2016jvj,Tachibana:2017syd,KunnawalkamElayavalli:2017hxo,Park:2018acg,Tachibana:2018yae}.

\section{\label{qhat_section}Jet Quenching}

Now that it has been demonstrated that the hadronic medium has an effect on the angular momentum distribution of shooting a single high-$p_T$ particle through it, we try to quantify this effect by calculating the transverse and longitudinal energy losses $\hat q$ and $\hat e$.

In the following we introduce a high-$p_T$ particle in the box simulating infinite matter described in Section \ref{smash} and analyse its first interaction with the medium averaged over many simulations. The reason for this is twofold. First, in a hadronic transport approach, inelastic collisions are not only possible but frequent (typically in the form of a string excitation, see Fig.~\ref{pip_xsec}), resulting in the loss of the original particle and thus making it impossible to continue to follow it. Second, even in the cases where the first interaction is elastic, it will typically not be in the form usually modeled at higher energies, where the medium is assumed to perform many relatively small kicks to the particle; in this case, a single collision can very strongly affect the momentum of the high-$p_T$ particle, and as such its final longitudinal momentum can vary a lot, making it difficult to compare subsequent collisions. By only studying the first collision the control parameters are kept fixed. 

In the QGP phase, $\hat q$ is typically estimated from kinetic theory \cite{Baier:2008js},
\begin{equation}
\hat q = \rho \int q_\perp^2 \frac{d\sigma}{dq_\perp^2} dq_\perp^2,
\label{qhat_kinetic}
\end{equation}
where $\rho$ is the density of the system, $q_\perp$ is the transverse momentum transfer and $d\sigma /dq_\perp^2$ is the differential cross-section of the particle with the medium. Since in SMASH the cross-section only depends on the properties of the incoming particles, the quantity $d\sigma /dq_\perp^2$ in this definition is consistent with zero, leading to the conclusion that Eq. \eqref{qhat_kinetic} should not be used in this context. This does not however mean that it is not possible to describe the energy loss in the hadronic medium. At the most fundamental level, the transverse and longitudinal energy losses are simply
\begin{equation}
\hat q = \frac{\langle q_\perp^2 \rangle_L}{L}, \qquad \hat e = \frac{\langle q_\parallel^2 \rangle_L}{L},
\label{qhat}
\end{equation}
where $\langle q_\perp^2 \rangle_L$ and $\langle q_\parallel^2 \rangle_L$ are the average cumulative transverse and longitudinal momentum change suffered by a propagating particle in a medium over a length $L$. As previously mentioned, this typically assumes that the medium acts through many small elastic kicks on the particle; this is not quite valid in the case of the hadron gas, as we rather usually observe one very large (and likely inelastic) momentum influx. To account for these differences in the description, we propose the following definitions for similar quantities denoted as $\tilde q$ and $\tilde e$,
\begin{equation}
\tilde q = \frac{\langle q_\perp^2 \rangle}{\lambda_{mfp}}, \qquad \tilde e = \frac{\langle q_\parallel^2 \rangle}{\lambda_{mfp}},
\label{qtilde}
\end{equation}
where $\langle q_\perp^2 \rangle$ and $\langle q_\parallel^2 \rangle$ are now the average transverse and longitudinal momentum of the first collision with the medium, and $\lambda_{mfp}$ is the mean free path of the high-$p_T$ particle before this first interaction.

\begin{figure}
  \includegraphics[width=85mm]{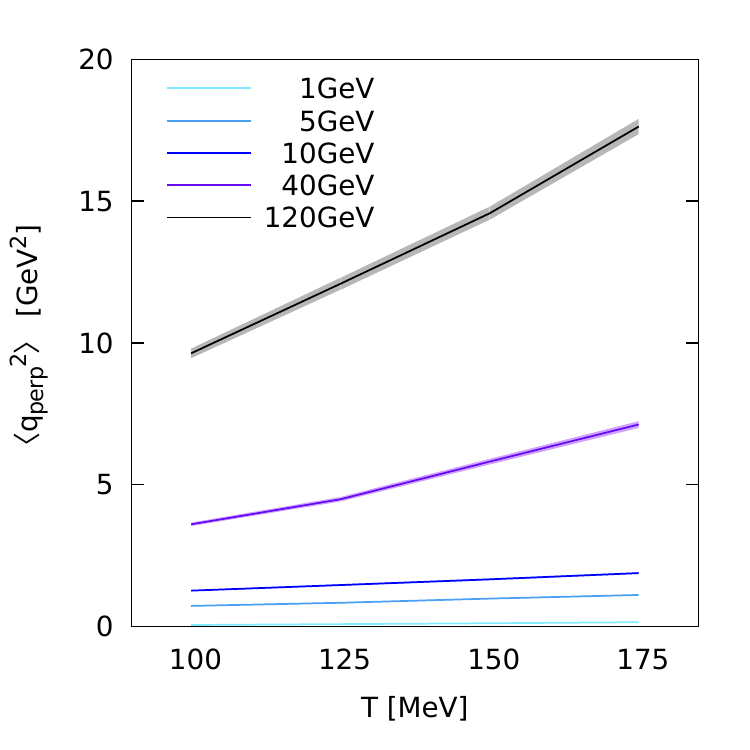}
  \hspace{0mm}
  \includegraphics[width=85mm]{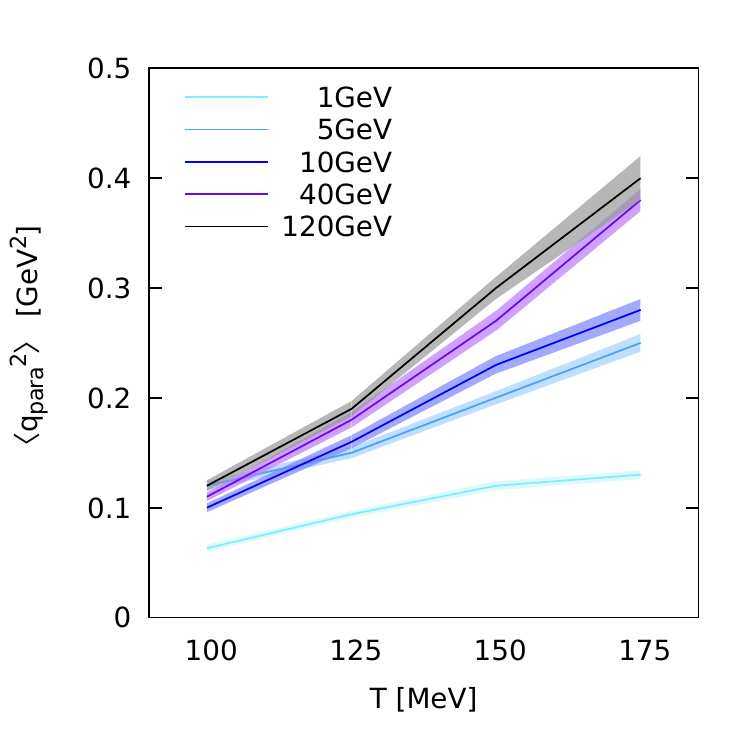}
  \caption{Transverse (top) and longitudinal (bottom) momentum vs temperature for various high-$p_T$ particle momenta.}
  \label{qperp_qpara}
\end{figure}

\begin{figure}
  \includegraphics[width=85mm]{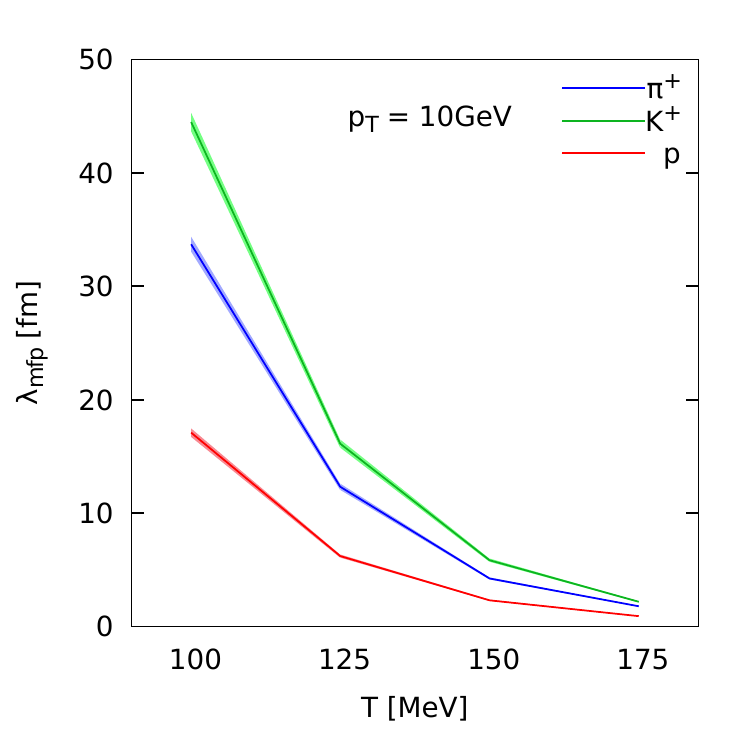}
  \caption{Mean free path vs temperature for various high-$p_T$ particle species.}
  \label{lambda_mfp}
\end{figure}

The top panel of Fig.~\ref{qperp_qpara} shows the effect of varying the energy of the high-$p_T$ particle on $\langle q_\perp^2 \rangle$, which is shown to increase with temperature at every energy; moreover, the effect of  temperature appears to be markedly more important as the energy of the particle increases. The bottom panel shows a similar picture for the case of $\langle q_\parallel^2 \rangle$, where we also see it increase with temperature at every energy; note however that in this case there is already a pretty strong temperature dependence even at low beam energies. Fig.~\ref{lambda_mfp} shows that the choice of species for the high-$p_T$ particle has a large impact on its mean free path, which mainly comes from the fact that cross-sections depend quite strongly on the type of particle. Although not shown here, the dependence of the transverse and longitudinal momenta are very similar for each species pointing to a main kinematic effect. In fact, the longitudinal momentum transfer is much smaller than the transverse one due to momentum conservation. Since the high momentum probe has initially rather high momentum, there has to be a significant amount maintained in the longitudinal direction, while any re-distribution into the transverse plane implies a large change since the transverse momentum is zero before the scattering. The mean free path on the other hand is rather insensitive to the energy of the particle as expected from the invariance of the string fragmentation versus the particle species and from the fact that the average cross-section only slightly varies with the energy of the particle (see for example Fig.~\ref{pip_xsec}). 

\begin{figure*}
  \includegraphics[width=85mm]{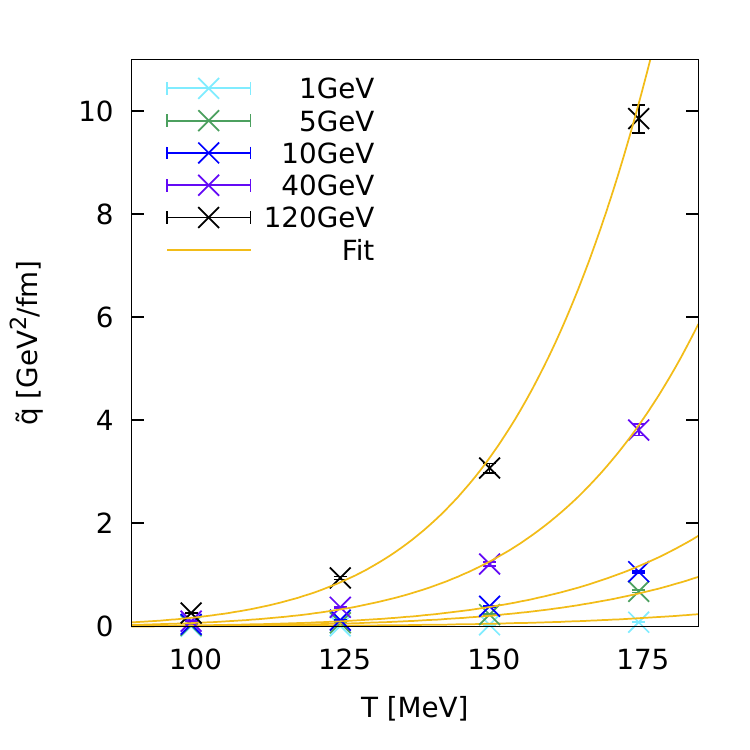}
  \includegraphics[width=85mm]{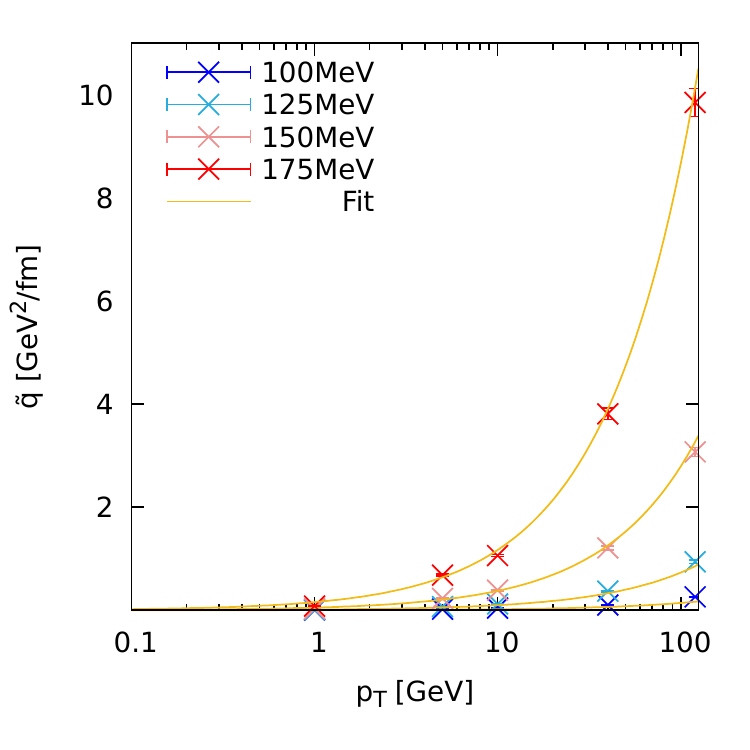}
  \includegraphics[width=85mm]{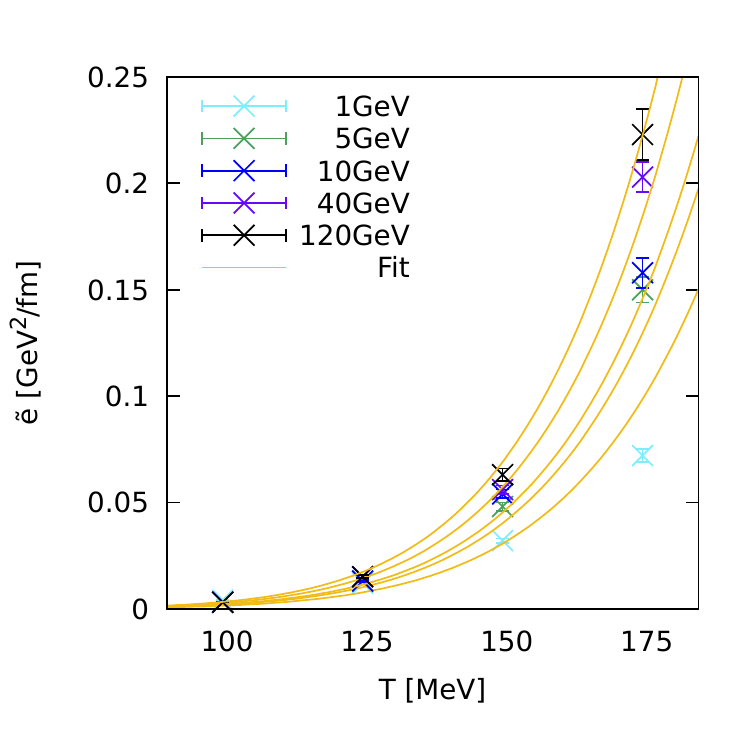}
  \includegraphics[width=85mm]{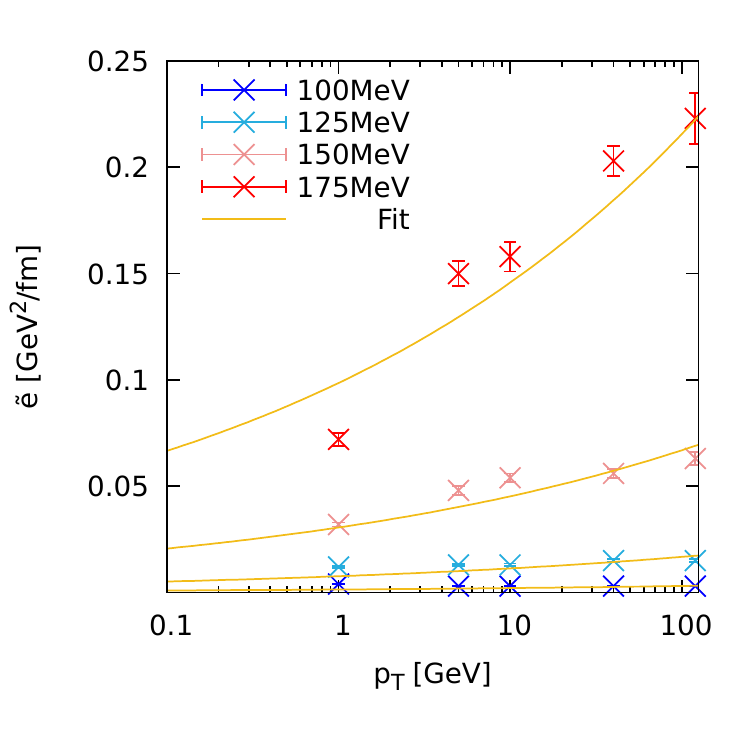}
  \caption{Medium modification factor $\tilde q$ as a function of temperature and high-$p_T$ pion momentum. Lines show the fit of parametrizations \eqref{param_qtilde} and \eqref{param_etilde}.}
  \label{qtilde_etilde_vs_params}
\end{figure*}

As a final note, we explore the temperature and particle energy dependence of $\tilde q$ and $\tilde e$ for the specific case of a pion high-$p_T$ particle (Fig.~\ref{qtilde_etilde_vs_params}). As one readily sees, we observe that both $\tilde q$ and $\tilde e$ increase significantly both with temperature and particle energy. This parameter space exploration allows us to introduce the following parametrizations (also shown on the figures) to estimate the value of the hadronic transverse and longitudinal energy losses at temperatures between 100 and 175 MeV, for particles with momenta between 1 and 120 GeV:
\begin{align}
\tilde{q}_\pi (p_T, T) &= 5.14 \cdot 10^{-18} \frac{\text{GeV}^2}{\text{fm}} \Big( \frac{p_T}{\text{GeV}}\Big)^{0.87} \Big( \frac{T}{\text{MeV}} \Big)^{7.35}, \label{param_qtilde}\\
\tilde{e}_\pi (p_T, T) &= 9.31 \cdot 10^{-19} \frac{\text{GeV}^2}{\text{fm}} \Big( \frac{p_T}{\text{GeV}}\Big)^{0.17} \Big( \frac{T}{\text{MeV}} \Big)^{7.59}. \label{param_etilde}
\end{align}
We observe that while both of these quantities depend on the momentum of the high-energy particle, the dependence is much stronger in the case of the transverse coefficient. These parametrizations can be directly employed in future studies of hadronic jet quenching. 

\section{\label{conclusion}Summary and Discussion}

We showed in Section \ref{shapes} that shooting a high-$p_T$ particle through a hadron gas does result in a broadening of the angular momentum distribution, or jet shape. As previously mentioned, this is a similar situation as to what should happen in the late stages of a typical heavy ion collision; the main difference is here the absence of flow. It is not trivial to predict how flow would affect these distributions at all energies. On the one hand, such an outwards movement of the particles will lead to a faster cooling and eventual freeze-out of the sphere; on the other, depending on the momentum of the high-$p_T$ particle, there are some cases in which lower $p_T$ particles would have a larger average cross-section with the medium since they would now be more likely to be (at least partially) comoving (see Fig.~\ref{pip_xsec} at $\sqrt s = 10$ GeV or lower, for example). Thus, while in the high-$p_T$ limit our results should represent a maximum as to what effect can be expected, the situation is not so clear at lower momentum.

We attempted to quantify those results through the calculation of the transverse and longitudinal energy losses in Section~\ref{qhat_section}. Although our results show that the kinetic expression for $\hat q$ is insufficient to obtain a physical result using our current calculation of the hadron gas, we proposed alternative definitions for the hadronic medium energy loss in the form of $\tilde q$ and $\tilde e$. Although it is not yet completely clear that the hat and tilde definitions are equivalent, we attempt a comparison between numerical values. In the QGP, a cross-model study has recently found that $\hat q = 1.9 \pm 0.7$ GeV$^2$/fm at T=470 MeV and $\hat q = 1.2 \pm 0.3$ GeV$^2$/fm at T=370 MeV \cite{Burke:2013yra} for a 10 GeV quark traveling through the medium. Our approach shows that for a pion with the same energy in a relatively hot hadronic medium at 150 MeV (i.e. just below the phase transition), we obtain from \eqref{param_qtilde} that $\tilde q = 0.38$ GeV$^2$/fm. This result, while as expected below its QGP counterpart, does remain significant. 

While this is beyond the scope of this exploratory study, the logical next step to verifying whether hadronic considerations are relevant to the jet quenching framework would be to perform full heavy ion simulations, complete with jets that first go through a QGP phase and then move on to cross a hadronic afterburner, for example within the JETSCAPE framework \cite{Putschke:2019yrg}.

\begin{acknowledgments}
The authors would like to thank Rainer Fries, Sangyong Jeon and J\"orn Putschke for extended discussions at the EMMI Rapid reaction task force on "The space-time structure of jet quenching: Theory and Experiment" in August 2019. This work was initiated during a visit in the context of the PPP exchange project with McGill University supported by the DAAD funded by BMBF (Project-ID 57314610). This work was supported by the Helmholtz International
Center for the Facility for Antiproton and Ion Research (HIC for FAIR) within the
framework of the Landes-Offensive zur Entwicklung Wissenschaftlich-Oekonomischer Exzellenz (LOEWE) program launched by the State of Hesse. Computational resources have been provided by the Center
for Scientific Computing (CSC) at the Goethe-University of Frankfurt. DP was supported by a grant from the Trond Mohn Foundation (project no. BFS2018REK01).
\end{acknowledgments}

\end{document}